%% file: 8322main.tex
\documentclass{aa}
\usepackage{txfonts,mathptmx}
\usepackage{natbib}
\bibpunct[, ]{(}{)}{;}{a}{}{,}
\usepackage{graphicx,epsf,textcomp,pstricks,psfrag}

\begin{document}

%-----------------------------------------------------------------------------------------
% Authors' Macros
%-----------------------------------------------------------------------------------------

\newcommand{\ie}{\textit{i.e.}\ }
\newcommand{\eg}{\textit{e.g.}\ }
\newcommand{\formula}{\begin{equation}}
\newcommand{\formulaend}{\end{equation}}
\newcommand{\vecbold}[1]{{\bf #1}}
\newcommand{\inv}[1]{\frac{1}{#1}}
\newcommand{\bytwo}[1]{\frac{#1}{2}}

\newcommand{\deriv}[2]{\frac{\mathrm{d} #1}{\mathrm{d} #2}}

\newcommand{\kilogram}{\mathrm{kg}}
\newcommand{\meter}{\mathrm{m}}
\newcommand{\cm}{\mathrm{cm}}
\newcommand{\mm}{\mathrm{mm}}
\newcommand{\micron}{\mu\mathrm{m}}
\newcommand{\AU}{\mathrm{AU}}
\newcommand{\kelvin}{\mathrm{K}}
\newcommand{\watt}{\mathrm{W}}
\newcommand{\parsec}{\mathrm{pc}}

\newcommand{\units}[2]{\left( \frac{#1}{#2} \right)}

\newcommand{\rout}{r_\mathrm{out}}

\newcommand{\gas}{\mathrm{g}}

\newcommand{\lp}{\, + \,}

\newcommand{\Terma}{\Upsilon_\mathrm{hc}}
\newcommand{\Termb}{\Upsilon_\mathrm{rad}}
\newcommand{\Termc}{\Upsilon_\mathrm{gas}}

\newcommand{\factorchia}{Z(\Terma)}
\newcommand{\factorchib}{Z(\Termb)}

\newcommand{\factorbeta}{B}
\newcommand{\factoreta}{E}

\newcommand{\stabilitydist}{r_\mathrm{stab}(s)}

%-----------------------------------------------------------------------------------------

\title{Effects of photophoresis on the evolution\\ of transitional circumstellar disks}

\bigskip
\author{Fabian Herrmann \inst{1}
        \and
        Alexander V. Krivov\inst{1}
       }
\offprints{Fabian Herrmann, \email{fabian@astro.uni-jena.de}
          }
\institute{Astrophysikalisches Institut, Friedrich-Schiller-Universit\"at Jena,
           Schillerg\"a{\ss}chen~ 2--3, 07745 Jena, Germany
          }

\date{Received July 20, 2007, accepted 1 October 2007}

\abstract
{
Although known for almost a century, the photophoretic force has only recently been 
considered in astrophysical context for the first time. In our work, we have examined the effect 
of photophoresis, acting together with stellar gravity, radiation pressure, and gas drag, on the 
evolution of solids in transitional circumstellar disks. We have applied our 
calculations to four different systems: the disks of HR 4796A and HD 141569A, which are 
several Myr-old AB-type stars, and two hypothetical systems that correspond to the solar
nebula after disk dispersal has progressed sufficiently for the disk to become optically thin. 
Our results suggest that solid objects migrate inward or outward, until they reach 
a certain size-dependent stability distance from the star. The larger the bodies, the
closer to the star they tend to accumulate.
Photophoresis increases the stability radii, moving objects to larger distances.
What is more, photophoresis may cause formation of a belt of objects, but only
in a certain range of sizes and only around low-luminosity stars.
The effects of photophoresis are noticeable in the size range from several micrometers to
several centimeters (for older transitional disks) or even several meters
(for younger, more gaseous, ones). We argue that due to gas damping, rotation does not substantially 
inhibit photophoresis.


}

\authorrunning{Herrmann \& Krivov}
\titlerunning{Photophoresis in transitional disks}
\maketitle

%-----------------------------------------------------------------------------------------
\section{Introduction}
%-----------------------------------------------------------------------------------------

The standard (core accretion) scenario of planet formation implies
continuous growth of solids in a protoplanetary disk around a young star
\citep[see, \eg\ ][]{safronov,wetherill-1980,shu_etal_87,lissauer_93,strom_edwards_93,
weidenschilling_cuzzi,papaloizou_etal_99,blum_wurm_2000,wurm_etal_06,henning_etal_06,meyer_etal_06}.
Micron-sized dust grows step by step to larger bodies, until asteroid-sized planetesimals are reached.
At sizes well below 1~km,  gravitational interactions between the objects play a minor role
and the process is largely determined by interactions of solids with the ambient gas in the disk.
Gas causes sedimentation, mixing, radial drift
\citep{weidenschilling_77,kley_etal_93,tak_arty,brauer_etal_07}
and other effects that all dictate the spatial, size, and velocity distributions
of grains and therefore the conditions for their growth.
As the disk evolves, both gas and dust are gradually removed from the systems.
In this process, the gas-to-dust ratio reduces from $\sim 100$ in the initial
protoplanetary disk stage, at the ages on the order of $1$~Myr,
to vanishing values of $\ll 1$ at the debris disk stage
after $\sim 10$~Myr
\citep{lawson_etal_04,haisch_etal_05,hollenbach_etal_05,
takeuchi_etal_05,jayawardhana_etal_06,balog_etal_07,currie-et-al-2007}.

Transition from gas- and dust-rich, optically-thick protoplanetary disks
to nearly gas-free, optically-thin debris disks is currently in the focus of
interest of both observational and theoretical effort
\citep{wieneke_clayton_83, morfill_83, morfill_88,
weidenschilling_cuzzi,strom_etal_93,zuckerman_becklin_93,simon_prato_95,meyer_beckwith_2000,
ardila,calvet_etal_05,hueso_guillot_05,augereau_06,bouwman_etal_06,eisner_etal_06}.
Examples of transitional objects are TW Hya, HR 4796A and HD 141569A, while the
well-known system $\beta$ Pic already enters into the realm of almost
gas-free debris disks \citep{thebault-augereau-2005}.
Resolved transitional disks exhibit radial structure in the form of rings and
gaps, often alternating, and there is a lot of debate whether this
structure is caused by gravity of hidden planets
\citep{augereau-et-al-1999a,wyatt-et-al-1999,telesco}
or by interaction of solids with the ambient gas component
\citep{klahr-lin-2000,tak_arty,besla_wu_07}.
The first possibility is supported by the fact that the amount of gas 
remaining in these systems is probably no longer sufficient to form Jupiter's
gas envelope \citep{chen-kamp-2004} and so, the planet formation process must
already be finished. The second hypothesis is substantiated by 
simulations based on the observational estimates of the gas
contents. These show that gas drag, acting together with other forces~--
stellar gravity and radiation pressure~--- could
result in segregation of 
different-sized solids and radial fractionation of dust.

In this paper, we study the motion of different-sized solids under the combined
action of stellar gravity, radiation pressure, and gas drag, to which we add
photophoresis, an additional radial force that
acts on particles in a gas disk which are exposed to the radiation field of the central 
star.
Photophoresis, one of the lesser known forces of physics, has been first described by 
\cite{ehrenhaft}. The force is caused by the 
following process. If an object is embedded in a thin gas and is exposed to an
anisotropic radiation field (\ie of a star), a gas molecule that becomes accommodated by 
the object's surface  and rejected again will depart 
from the illuminated -- and therefore warmer -- side on average with a greater velocity 
then from the dark, colder one. Net momentum is transferred to the object,
accelerating it away from the light source \citep[note that for very small
particles, the force can also be attractive -- see][]{tehranian_etal_2001}.
Although known for almost a hundred years and
successfully used in technical applications, such as the construction of optical traps 
\citep[][]{steinbach_etal_04},
it has only recently been analyzed in astrophysical context.
The first attempt to examine its influence on planetary formation was made by 
\cite{krausswurm}. In their subsequent paper \citep[][]{krausswurm_cai}, they investigated
how photophoresis might affect the formation of chondrules and the survival of 
Calcium/Aluminum-rich inclusions. Finally, \cite{krausswurm_06}
analyzed the effects imposed by photophoresis on the inner rim of a dusty protoplanetary disk.
They assumed the disk to stay optically thick at all times, thus allowing only 
its inner edge to be influenced by photophoresis. Using an exponential law to
describe particle growth and the $\alpha$ 1+1D turbulent model \citep[][]{shakura_sunyaev_73} 
used in \cite{alibert_05} and \cite{papaloizou_terquem_99}
for the evolution of the gas disk, Krauss and Wurm
computed the motion of the inner edge. This is identical to the motion of the smallest 
particles, since, as will be shown subsequently,
for solar luminosity systems, the
velocity of outward motion of an object increases with its size in a certain
size range (compare Fig. \ref{radvel}). In essence, they showed that 
the dust disk's edge first moves outward, pushed by photophoresis. After a 
few million years of time, it shrinks towards the star again, because gas pressure declines as 
disk dispersal proceeds.

This paper focuses on transitional disks. In these systems, the conditions are quite
favorable for photophoresis to be efficient.
Transitional disks are already optically thin, so that the particles are exposed
to the strong radiation field. On the other hand, they
are still sufficiently gas-rich.
We would like to check whether, and to which 
extent, photophoresis may affect the radial migration of solids.

We use a single-body dynamics approach to investigate the behavior of a two-dimensional swarm of
particles, the sizes of which are distributed in an interval ranging from
several micrometers to several meters.
The size of an individual particle, though, is being kept constant.
We develop a simple theory for the long-term behavior of a distribution of solids
experiencing gravity, radiation pressure, gas drag and photophoresis. Methodically, our study is 
similar to \cite{tak_arty}, whose computations we generalize to
the presence of photophoresis. To solve the equation of motion, we employ a
modification of the semianalytic scheme used by \cite{weidenschilling_77}.
The results are applied to circumstellar disks with different gas density
around stars of different luminosity.

In Sect. 2, we discuss the astrophysical status and parameters of the
systems chosen for this study, as well as the formulas used for gas density and temperature in 
the disks. Section 3 provides formulas for different perturbing forces, including
the photophoretic one.
Section 4 deals with the equation of motion.
In Sect. 5, we analyze the properties of the particles' radial
motion and derive a formula for the radii of stable orbits. We also estimate
the size ranges in which photophoresis should be taken into account.
In Sect. 6, we examine the problem of particle rotation in order to check
whether it might nullify the effect of photophoresis.
In Sect. 7, the results are summarized and ideas for future research are
presented.

%-----------------------------------------------------------------------------------------
\section{The systems}
%-----------------------------------------------------------------------------------------

\begin{table*}
\caption{\label{model_pars} Parameters of gas disk models.}
\begin{tabular}{llllrr}
\hline
System & Example & $\rho_0 \, [10^{-10} \, \kilogram \, \meter^{-3}]$ &
Density exponent $q$ &
Luminosity $L_\star / L_\odot$ &
Mass $M_\star / M_\odot$ \\
\hline
lGlL & `older' SN     &  3.12 & -2.75 &  1.0 & 1.0 \\
hGlL & `younger' SN   &  156  & -2.75 &  1.0 & 1.0 \\
lGhL & HR 4796A$^*$   &  3.12 & -2.75 & 21.0 & 2.5 \\
hGhL & HD 141569A$^*$ &  156  & -2.25 & 22.4 & 2.3\\
\hline\\
\end{tabular}

{\small $^*$Sources: \cite{tak_arty}, their Table~1}
\end{table*}

Following \cite{tak_arty}, we choose two transitional disks of particular 
interest for our study: HR 4796A and HD 141569A. 
We use these two systems in our work as model environments for
the forces we wish to explore, as their properties are quite well known and
reasonably accurate models exist for the physical conditions (gas density
and temperature) in their disks. 

The A0-type star HR 4796A,
located at a distance of $67.1^{+3.5}_{-3.4} \,
\parsec$ from the sun (\textit{Hipparcos} data),
is a member of the TW-Hydrae-Association
\citep[TWA in the following, see][]{kastner_etal_01}.
According to current knowledge, this is the young star association closest to our sun
\citep[][]{zuckerman_song_04}. The age
of TWA can be estimated to be approximately 10~Myr. Since it has
used up its molecular cloud completely, the only gas left in in the
association is bound in the circumstellar disks.
The stellar parameters of HR 4796A are estimated to be
$M_\star = 2.5 \, M_\odot, \; L_\star = 21.0 \, L_\odot$
\citep[][]{koerner,jaya,telesco}.
IRAS discovered that it
emits 0.5 \% of its entire radiative power in the infrared part of the
spectrum, which points to a high dust density in its neighborhood -- in
fact, HR 4796A is the dust-richest star in the \textit{Bright Star Catalog}.
\cite{stauffer} estimate its age to be $8 \pm 2 \, \mathrm{Myr}$,
which makes it slightly younger than $\beta$ Pic (12 Myr). The structure of the
dust disk around HR 4796A is highly complicated. First resolved images in
thermal infrared were
obtained by \cite{koerner} and \cite{jaya}. \cite{wahhaj_etal_05}
conducted elaborate studies of the disk's structure, using data from MIRLIN
at the Keck II telescope as well as $350 \, \micron$ measurements from the
Caltech Submillimeter Observatory and Hubble Space Telescope scattered light
images. Their studies point to a disk composed of an inner, exozodiacal dust
ring located at roughly $r = 4 \, \AU$ from the star and a wide,
two-component outer dust belt consisting of a broad ring of $\sim 7 \,
\micron$ grains stretching from 45 to 125 AU, and a narrower structure
between 66 and 80 AU consisting of $\sim 50 \, \micron$ grains. While the
exozodiacal dust may be the product of an asteroid-type belt, the outer belt
can be explained most naturally by the
assumption that the grains are emitted by an exo-Kuiper swarm of planetesimals.
Since the smallest ones are diffused rapidly by radiation pressure, they form a
broad belt, while the larger ones stay closer to their area of origin.
Also, asymmetric structures have been reported \citep{telesco}.
They may be caused by secular perturbations
of one or several planets or by the stellar companion HR 4796B. Most
probably, the disk is perturbed by planetary bodies as well as by the B
component. In order to define the mass and number of planets, the orbit
of HR 4796B would have to be known with greater precision.

The second object, HD 141569A,
has the \textit{Hipparcos} distance of $99 \pm 10 \, \parsec$.
Its parameters are estimated as
$M_\star = 2.3 \, M_\odot, \; L_\star = 22.4 \, L_\odot$
\citep[][]{jura,jura_etal_98,vandenancker_etal_98}
It is underluminous for 
its spectral type B9.5
Ve, which is a common occurrence for young A-type stars, found also at HR
4796A, $\beta$ Pic and 49 Ceti
\citep[][]{jura_etal_98,lowrance_etal_2000}.
Being, like HR 4796A, part of
a multiple system with an M2 and an M4 component, the lower-mass companions
which have not yet reached the main sequence can be used to find an estimate
for the system's age of $5 \pm 3$ Myr \citep[][]{weinberger}.
As in the HR 4796A case, the disk itself has a complex morphology
 \citep[][]{weinberger_etal_99,augereau_etal_99,fisher_etal_2000,mouillet_etal_01},
consisting of two dust belts with radii of 200 and 325 AU, the centers of
which are shifted by 20 -- 30 AU in the direction of the system's semiminor axis.
Between them a dust-free gap can be found at $\sim$ 250 AU. Also, inside of
150 AU the disk's luminosity declines to the level of background noise,
which points to a strong depletion of dust in
the system's inner region. The outer ring shows a tightly-wrapped spiral
structure, which, according to the numerical calculations of
\cite{aug_papal}, can be produced by the gravitative perturbation of HD
141569B and C. The dust gap at 250 AU may be caused by a planet of
approximately Jovian mass, but it is not yet clear whether gas giants can
form within a few million years at a distance of several hundred
AU from the star \citep{wyatt_05}.

Both disks~-- HR 4796A and HD 141569A~-- surround luminous, early-type stars.
Since we wish to explore the dependence of the particle dynamics on the stellar luminosity as well,
we add two more, hypothetical systems with gas densities equal to those of HR 4796A and HD 
141569A, but around a star of solar luminosity. They can be viewed as representations of the 
solar nebula at different stages of dissipation 
\citep[][]{hollenbach_etal_94,hollenbach}.

For gas density and temperature in all four model systems, we use standard power-law 
approximations \citep[][]{hayashi}:
\formula
\frac{T}{1 \, \kelvin} = 278 \; \units{L_\star}{L_\odot}^{1/4} \; \units{r}{1 \, \AU}^{-1/2}
\label{Temp}
\formulaend
and
\formula
\frac{\rho_\gas}{1 \, \kilogram \, \meter^{-3}} = \rho_0 \; \units{r}{1 \, \AU}^{q} \, ,
\label{rho_gas}
\formulaend
where $\rho_0$ and $q$ are constants that vary from one system to another.
To keep the treatment simple, we assume the exponent $q$ to be constant throughout
each disk and do not use
a `break-off function' to describe the density drop beyond a certain radius 
$\rout$ \citep{tak_arty}.
As will be shown subsequently, the outer part of the radial profile does not really 
matter, because photophoresis affects only the inner
parts of the disk.

Altogether, our model contains three basic parameters:
the gas density $\rho_0$ at $r = 1 \, \AU$, the gas density exponent $q$ which
controls the size of the gas disk, and the stellar luminosity $L_\star$.
Our four model systems essentially explore the photophoretic effect in
the `parameter rectangle' luminosity -- gas density:
\begin{itemize}
\item
{\em lLhG} (low luminosity, high gas content, corresponding to the solar nebula at
an earlier stage),
\item
{\em lLlG} (low luminosity, low gas content, solar nebula at a later stage),
\item
{\em hLhG} (high luminosity, high gas content, HD 141569A), and
\item
{\em hLlG} (high luminosity, low gas content, HR 4796A).
\end{itemize}
The parameters of all four systems are listed in Table \ref{model_pars}.

%-----------------------------------------------------------------------------------------
\section{Forces}
%-----------------------------------------------------------------------------------------

In circumstellar nebulae, solid bodies experience a number of perturbing
forces in addition to gravity, causing them to move along non-Keplerian
orbits. In a first approximation, we can assume that they describe circular
orbits, the radii of which shrink or grow in the course of time, depending
on the size of the body and its distance from the star.
We now introduce the different forces, presenting an analytic expression for
each.

\subsection{Photophoresis}

While \cite{rohatschek_96} found a semi-empirical formula for the photophoretic force at all gas pressures,
\cite{beresnev} derived an analytic expression for spherical, nonrotating objects with a homogenous 
surface using an elaborate theoretical approach
starting from the molecular velocity distribution function \citep[compare also][]{tehranian_etal_2001}. 
The acceleration due to the photophoretic force they found is:
\begin{eqnarray}
a_\mathrm{phot} &=&
\frac{I \, J_1}{4 \, s \, \rho_\mathrm{bulk}} \;
\sqrt{\frac{\pi \, \mu_\mathrm{g} m_\mathrm{H}}{2 k T}} \;
\nonumber\\
&\times&
\frac{\alpha_\mathrm{E} \Psi_1}{\alpha_\mathrm{E} \, + \, 15 \Lambda Kn \, (1
- \alpha_\mathrm{E})/4 \, + \, \alpha_\mathrm{E} \Lambda \Psi_2} \, .
\label{phot_beschl_allkn}
\end{eqnarray}
Here, $s$ is the particle radius, $I$ the radiation intensity, $J_1$ the
asymmetry parameter that describes the accommodation of gas molecules to the particle's surface 
and light absorption. Assuming complete absorption and an accommodation probability
of 100 \%, we set $J_1 = 0.5$. The mean molecular weight of a gas of solar
composition is denoted by $\mu_\mathrm{g} = 2.34$ and
the mass of a hydrogen atom by $m_\mathrm{H}$.
Further, $k$ is Boltzmann's constant, $T$ the gas
temperature and $\rho_\mathrm{bulk}$ the density of the solid material. For icy aggregates,
$\rho_\mathrm{bulk} \approx 1000 \, \kilogram \, \meter^{-3}$ is a good approximation. 
The energy accommodation coefficient $\alpha_\mathrm{E}$ is the fraction of
molecules in contact with the surface that accommodate to the local
temperature, which enables them to contribute to photophoresis. In our work
we assume complete accommodation, and thus set $\alpha_\mathrm{E} = 1$. The
heat exchange parameter $\Lambda$, which describes the particle's thermal relaxation
properties, is determined as
\formula
\Lambda = \frac{k_\mathrm{th} + 4 \epsilon \sigma T^3 s}{k_\mathrm{gas}} \,
,
\formulaend
where $k_\mathrm{th}$ is the material's thermal conductivity, $\epsilon$ its
emissivity, $\sigma$ the Stefan-Boltzmann constant and $k_\mathrm{gas}$ the
thermal conductivity of the gas. Assuming black bodies, we set $\epsilon = 1$.
The quantities $\Psi_1$ and $\Psi_2$ are given by
\begin{eqnarray}
\Psi_1 = \frac{Kn}{Kn \, + \, (5 \pi / 18)} \; \left( 1 \, + \, \frac{2
\pi^{1/2} \, Kn}{5 \, Kn^2 \, + \, \pi^{1/2} \, Kn \, + \, \pi / 4} \right)
\, , \nonumber \\
\Psi_2 = \left( \frac{1}{2} \, + \, \frac{15}{4} \, Kn \right) \; \left( 1
\, - \, \frac{1.21 \, \pi^{1/2} \, Kn}{100 \, Kn^2 \, + \, \pi/4} \right) \,
,
\end{eqnarray}
and $Kn$ is the Knudsen number.
It is defined as $Kn \equiv L / s$,
where $L = 1 / (\sqrt(32) \, \pi \, n_\mathrm{gas} \, r_\mathrm{g}^2)$
is the molecule's mean free path,
with $r_\mathrm{g} \approx 10^{-10}$~m being the radius of gas molecules
and $n_\mathrm{gas}$ their number density.

\begin{figure}[t]
\centerline{
\epsfxsize = 0.95\columnwidth
\epsffile{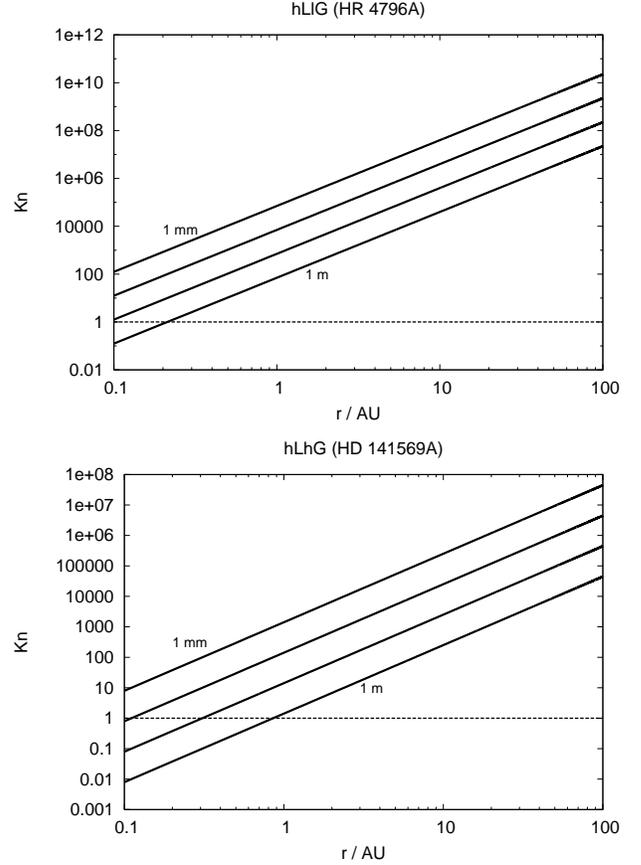}}
\caption{\label{meanfree}
 Knudsen numbers $Kn = L / s$ for systems with low (top) and high (bottom) gas content
as a function of distance from the star.
In each panel, four lines correspond to the particle radii of
1~mm, 1~cm, 10cm, and 1~m.
Since $L$ only depends on gas density, $Kn$ does not vary with stellar luminosity.
}
\end{figure}

Expression Eq.~(\ref{phot_beschl_allkn})
for the photophoretic acceleration is valid for all
Knudsen numbers.
In the case of high Knudsen numbers (mean free paths of molecules are
large compared to particle sizes), 
Eq.~(\ref{phot_beschl_allkn}) reduces to (see \cite{krausswurm}, their
Eq. 1, and \cite{beresnev}, their Eq. 26)
\formula
a_\mathrm{phot} = \frac{I p J_1}{4 \rho_\mathrm{bulk} \; (\Terma \lp \Termb \lp \Termc)} \, ,
\label{phot_beschl}
\formulaend
where
\begin{eqnarray}
\Terma = k_\mathrm{th} T \, , \label{A} \\
\Termb = 4 \sigma T^4 \epsilon \, s \, , \label{B} \\
\Termc = p \sqrt{2kT/\pi \mu_\mathrm{g} m_\mathrm{H}} \; s \, .
\label{C}
\end{eqnarray}
Here, $p = n_\mathrm{gas} k T$ is the gas pressure (assuming ideal gas).
In transitional disks, we can usually assume to be in the high Knudsen
number regime, except for large objects in the innermost parts of the disks
(Fig.~\ref{meanfree}).
The simplified expression (\ref{phot_beschl}) will allow us
to find useful approximate solutions for the particles' stability radii
(\ie\ the radii of stable circular orbits). For large particles ($s \ge 10
\, \cm$) and small distances from the star ($r \le 1 \, \AU$), the more general
expression (\ref{phot_beschl_allkn}) will be used in numerical calculations.

The terms $\Terma$, $\Termb$ and $\Termc$ are related to three different processes which reduce 
the photophoretic effect.
The first term, $\Terma$, corresponds to the
\textit{transport of thermal energy through the object} -- \ie heat conductivity, which reduces 
the temperature gradient, thus lowering the efficiency of photophoresis.
The second one, $\Termb$, describes
\textit{thermal radiation from the particles' surface}.
Since it is proportional to $T^4$, much more energy is radiated from the warm than from the 
dark side, which results
in a substantial reduction of the temperature gradient, and, therefore, of photophoresis. 
Obviously, this process gains efficiency with increasing temperature.
Finally,
the third term $\Termc$ describes \textit{heat conduction away from the object's surface into 
the  surrounding gas}.
Since it also shows a weak dependence on temperature ($\propto T^{1/2}$), 
it causes a reduction of temperature gradient too.

\begin{figure}[h]
\centerline{
\epsfxsize = 0.95\columnwidth
\epsffile{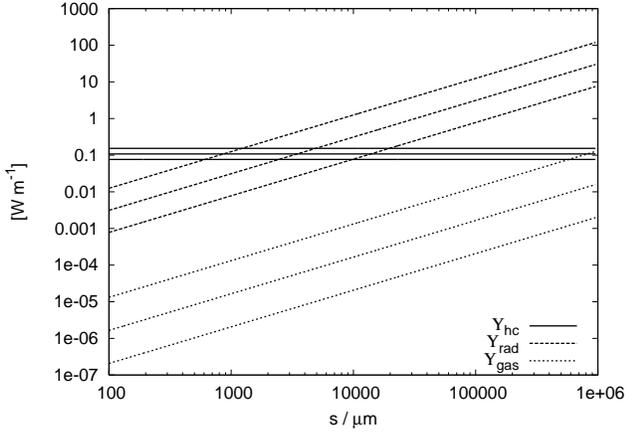}}
\caption{\label{comp}
Quantitative comparison of the three reduction processes
for hLhG (HD 141569A) at $r = 5, \, 10, \, 20 \, \AU$ (from above for each
set of lines). For the other systems the result looks qualitatively
similar, with $\Termc$ being in the case of the gas-poorer systems completely negligible
for all sizes $s$.}
\end{figure}

\cite{krausswurm} and \cite{krausswurm_cai} use only the first term $\Terma$ in the denominator 
of Eq.~(\ref{phot_beschl}), which is a good approximation for particles with 
$s \ll 1 \, \mm$ and only in the case of low luminosity (\ie solar-type) systems
(see Sect.~\ref{eq_distance}).
In order to analyze the relative importance of the three `countereffects' $\Terma$, $\Termb$ and
$\Termc$, we plot them for HD 141569A as functions of particle size (Fig. \ref{comp})
at three different distances.

Figure \ref{comp} shows that for fairly small particles ($s \le 1 \, \mm$), heat conduction ($\Terma$) 
is the main reduction process. At larger sizes ($s \ge 10 \, \cm$), radiation ($\Termb$) takes over. 
In the intermediate interval, both processes contribute to the reduction of photophoresis.
At yet larger sizes ($s \ge 1 \, \meter$), heat conduction from the surface into the surrounding gas 
($\Termc$) comes into play, approaching values similar to that of $\Terma$. As lower gas pressure 
dramatically reduces the importance of $\Termc$, it is completely negligible for all object radii 
in the case of the gas-poor systems hLlG (HR 4796A) and lLlG.

\subsection{Radiation pressure}

Since radiation pressure is proportional to $r^{-2}$, it can be taken into account by
introducing an `effective stellar mass' \citep{burns-et-al-1979}:
\formula
M_\mathrm{eff} = M_\star \, (1 - \beta) \, ,
\formulaend
where $\beta$ stands for the ratio of accelerations due to radiation pressure 
and gravity:
\begin{eqnarray}
\nonumber
\beta 
= \frac{a_\mathrm{rad}}{a_\mathrm{grav}}
= \frac{0.5738 \; Q_{\mathrm{pr}}}{\rho_\mathrm{bulk} [\mathrm{g \: cm}^{-3}] \, s[\mathrm{\mu m}]} \: 
\frac{L_\star / L_\odot}{M_\star / M_\odot}\\
\equiv \factorbeta \; \units{s}{1 \, \micron}^{-1} \, ,
\label{beta}
\end{eqnarray}
with $Q_{\mathrm{pr}}$ being the radiation pressure efficiency.
We will make use of the constant $B$ later.
The effective `photogravitational' acceleration is then
\formula
\vecbold{a}_\mathrm{grav, \, eff} = -G \frac{M_\mathrm{eff}}{r^3} \; \vecbold{r} \, .
\label{a_grav}
\formulaend
In our work, we simply set $Q_{\mathrm{pr}}=1$,
assuming the solids to be black bodies.
The reason is that generalization to non-black surfaces requires
a change not only in the emissivity, but also in the asymmetry parameter $J_1$.
The latter is not known for realistic materials, and its determination
requires complicated numerical procedures \citep[see, \eg\ ][]{mackowski}.
Analytical solutions are possible, but only in the cases of
high or low absorption coefficients \citep[see][]{arnold_lewittes_82}.

\subsection{Gas drag force}

\label{gasdrag}

Unlike the other three forces (gravity, radiation pressure, photophoresis),
gas drag is not a radial force. Since it results from the momentum
transferred to the body by molecules impinging on it as it moves through the
gas, the force vector is antiparallel to the relative velocity $\Delta \vecbold{v}$
of the dust grain
with respect to the gas.

To calculate gas drag, we first need a model for the motion of the
gas component of the circumstellar nebula. 
Since the pressure gradient supports gas against stellar gravity,
it travels on circular orbits with a sub-Keplerian speed
\formula
v_\gas = v_\mathrm{K} \, \sqrt{1 - \eta} \, ,
\label{gas_vel}
\formulaend
and angular velocity
\formula
\Omega_\mathrm{g} = \Omega_\mathrm{K} \, \sqrt{1 - \eta},
\formulaend
where
$v_\mathrm{K}  = \sqrt{G M_\star / r}$ is the Keplerian circular velocity,
$\Omega_\mathrm{K} = v_\mathrm{K} / r = \sqrt{G M_\star / r^3}$ the corresponding angular velocity,
and $\eta$ is the ratio of pressure gradient force to gravity:
\formula
\eta 
= - \inv{r \Omega_\mathrm{K}^2 \rho_\gas} \, \deriv{p}{r} \,
.
\label{eta}
\formulaend
The $\eta$ ratio can be rewritten as:
\begin{eqnarray}
\nonumber \eta &=& 1.1 \times 10^{-3} \, \left( \inv{2} - q \right) \: 
\units{\mu_\mathrm{g}}{2.34}^{-1} 
\, \units{L_\star}{L_\odot}^{1/4} \, \\
& \times & \, \units{M_\star}{M_\odot}^{-1} \, \units{r}{1 \, \AU}^{1/2} 
\equiv \factoreta \; \units{r}{1 \, \AU}^{1/2} \, .
\label{eta_lang}
\end{eqnarray}

If the motion of the particles is subsonic ($\Delta v \ll v_\mathrm{T}
\approx c_\mathrm{S}$ where $c_\mathrm{S}$ is the speed of sound -- this 
is always the case in the transitional disks), the gas drag acceleration is given by
\citep[see][]{tak_arty}:
\formula
\vecbold{a}_\mathrm{D} = - \frac{3 \rho_\mathrm{g}}{4 \rho_\mathrm{bulk} s} \: v_T
\: \Delta \vecbold{v} \, ,
\label{reib_vollst}
\formulaend
where
\formula
v_\mathrm{T}
= \frac{4}{3} \left( \frac{8kT}{\pi\mu_\mathrm{g} m_\mathrm{H}} \right)^\inv{2}
= \frac{4}{3} \times \langle v_\mathrm{therm} \rangle
\formulaend
is $4/3$ times the mean thermal velocity.

The reaction of particles to the gas drag force critically depends on their
size. Small objects adjust their velocity instantaneously, they are swept
along with the gas component. Large ones react only sluggishly to gas drag,
taking a longer time to change their velocity substantially. In order to
create a quantitative measure for the tendency of solids to be influenced by
gas drag, the \textit{stopping time} is introduced.  
We denote the time needed for a particle injected into the gas to be slowed down to $e^{-1}$ 
times its initial velocity by $t_\mathrm{s} = \Delta v \, / \, a_\mathrm{D}$. The 
dimensionless stopping parameter $T_\mathrm{s}$ is defined by:
\formula
T_\mathrm{s} \equiv t_\mathrm{s} \, \Omega_\mathrm{K} \approx \frac{4 \rho_\mathrm{bulk} s 
v_\mathrm{K}}{3 \rho_\gas r v_\mathrm{T}} \, .
\label{T_s}
\formulaend
The right-hand side approximation holds in the case of subsonic motion. It
renders the stopping time independent of the particle's momentary velocity.

%-----------------------------------------------------------------------------------------
\section{Equation of motion}
%-----------------------------------------------------------------------------------------

The equation of motion of a dust particle is:
\formula
\frac{\mathrm{d}^2 \, \vecbold{r}}{\mathrm{d} \, t^2} = \vecbold{a}_\mathrm{grav, \, eff} +
\vecbold{a}_\mathrm{phot} + \vecbold{a}_\mathrm{D} \, .
\label{bew_gl}
\formulaend

Being interested in the radial motion of the bodies, we now consider
the radial component of the equation of motion~(\ref{bew_gl}).
In the reference frame corotating with the gas, it takes the form
\formula
{\mathrm{d}^2 r \over \mathrm{d} t^2 } \, - \, \Omega_\mathrm{g}^2 \, r = \;
- a_\mathrm{grav, \, eff} + \, a_\mathrm{phot} \, .
\formulaend
The acceleration $\mathrm{d}^2 r / \mathrm{d} t^2 \equiv \Delta g$ of the particle
in the corotating frame can be interpreted as the `residual gravity'.
It computes to
\formula
\Delta g = - a_\mathrm{grav, \, eff} + a_\mathrm{phot} + \Omega_\mathrm{g}^2 r
= (\beta + \chi - \eta) \, \Omega_\mathrm{K}^2 \, r \,,
\formulaend
where $\chi \equiv a_\mathrm{phot} / a_\mathrm{grav}$ is the
photophoresis-to-gravity ratio.
Particles experience the inward-directed
residual acceleration $\eta \Omega_\mathrm{K}^2 r$ and the outward-directed
accelerations due to  photophoresis ($\chi \Omega_\mathrm{K}^2 r$)
and radiation pressure ($\beta \Omega_\mathrm{K}^2 r$).

Assuming now that the solid particle moves at a Keplerian circular speed,
we can write
\formula
{v_\mathrm{g}^2 \over r} = {v_\mathrm{K}^2 \over r} + \Delta g \, .
\formulaend
The relative velocity is then approximately
\formula
\Delta v = v_\mathrm{K} - v_\mathrm{g} \approx - \left( \frac{\Delta g}{2 \,
a_\mathrm{grav}} \right) \; v_\mathrm{K} \, .
\formulaend

We now use $\Delta g$ to derive an approximation for the
radial drift velocity of the particles. 
We can discern two limiting cases:
\begin{itemize}
\item {\em Small particles} are swept along with the gas. They stay
on circular orbits, while moving with sub-Keplerian velocity, which causes
them to experience the residual gravity pull $\Delta g$.
\item {\em Large particles} are decoupled from gas.
They move on Keplerian orbits, experiencing a head wind that
gradually reduces their angular momentum.
\end{itemize}
In the first case, the radial drift velocity 
$v_\mathrm{r} \equiv \mathrm{d} r / \mathrm{d} t$
computes to 
\citep[see][sect. 4.1, 4.2]{weidenschilling_77}:
\formula
v_\mathrm{r, \, small}
= t_\mathrm{s} \, \Delta g \, = T_\mathrm{s}
(\beta + \chi - \eta) \, r \Omega_\mathrm{K} \, .
\formulaend
In the second case, the orbit decays at a rate
\formula
v_\mathrm{r, \, large}
= - \frac{r}{t_\mathrm{s}} \, \frac{2}{v_\mathrm{K}} \, \Delta v
= \frac{r}{t_\mathrm{s}} \, \frac{\Delta g}{a_\mathrm{grav}} = \frac{(\beta
+ \chi - \eta) \, r \Omega_\mathrm{K}}{T_\mathrm{s}}
\, .
\formulaend
These two formulas can be combined into
\citep[see][Sect. 3.3]{tak_arty}:
\formula
v_\mathrm{r} = \frac{\beta + \chi - \eta}{T_s + T_s^{-1}} \, v_K \, .
\label{vau_r}
\formulaend

\begin{figure}[h]
\centerline{
\epsfxsize = 0.95\columnwidth
\epsffile{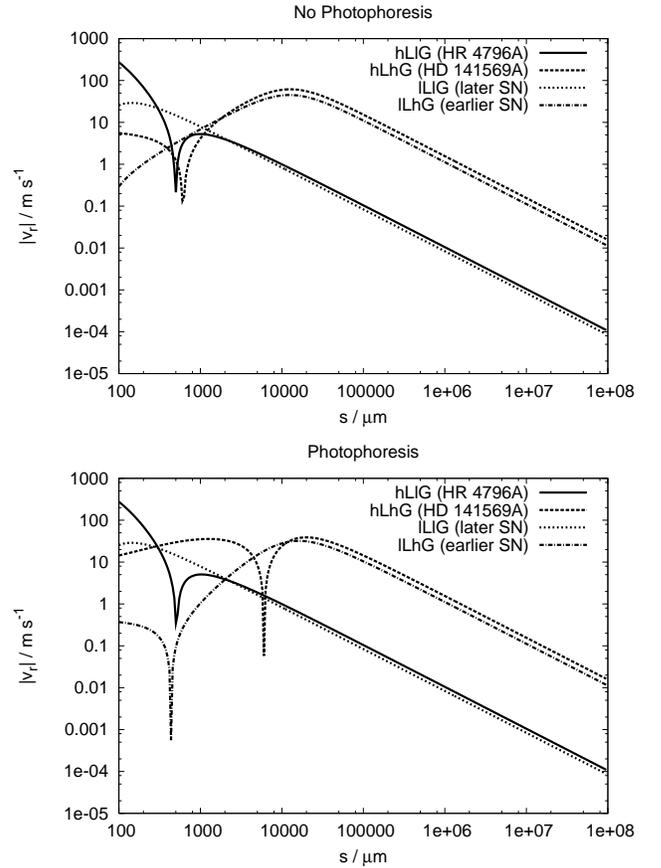}}
\caption{\label{radvel} The absolute values of the radial velocities at $r = 10 \, \AU$. The 
minima correspond to the sizes for which $v_\mathrm{r} = 0$.}
\end{figure}

In Fig. \ref{radvel}, we plot the absolute value of $v_\mathrm{r}$ over $s$ for
$r = 10 \, \AU$, both with and without photophoresis.
The points where $v_\mathrm{r}=0$ are
shifted to larger sizes by photophoresis. Note that for larger sizes,
radial velocities depend almost exclusively on gas density since for them gas drag
($\eta$ in Eq.~\ref{vau_r}) becomes the main driving force of radial
migration.

In what follows, differential equation (\ref{vau_r}) for $r(t)$
will be studied analytically and solved numerically.
In the realm of large radial velocities (or equivalently, small particles),
$r(t)$ should be computed from the `exact' equation of motion (\ref{bew_gl}).
However, our numerical tests have shown Eq. (\ref{vau_r}) to be accurate enough
for all particle sizes larger than $\sim 100 \, \micron$.

%-----------------------------------------------------------------------------------------
\section{Results}
%-----------------------------------------------------------------------------------------

\subsection{Radial motion}

We used Eq.~(\ref{vau_r}) to calculate the radial drift
of different-sized particles.
What happens, in short, is that the particles migrate inward or outward, until they reach 
stable circular orbits on which gravity, centripetal force (in the particle's own
inertial system), photophoresis and radiation pressure balance each other, thus permitting 
circular motion ($v_\mathrm{r} = 0$), while at the same time $\Delta v = 0$, \ie the particle 
travels at the same speed as the gas.

\begin{figure}[h]
\centerline{
\epsfxsize = 0.95\columnwidth
\epsffile{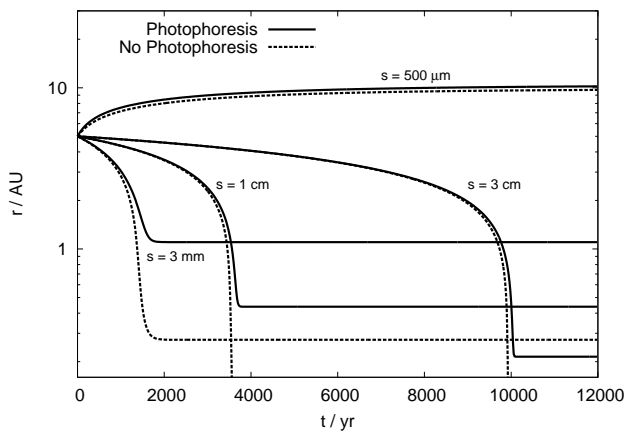}}
\caption{\label{rt_bsp_hr} Distance as a function of time for particles of four
different radii in the hLlG system (HR 4796A). Dashed lines: without photophoresis,
solid: with photophoresis. The lower border of the plot corresponds
to the evaporation limit $r_\mathrm{evap} = 0.16 \, \AU$
(defined by black-body equilibrium temperature $T = 1500 \, \kelvin$).}
\end{figure}

As an example, Fig. \ref{rt_bsp_hr} shows the $r(t)$-curves for particles of four
different sizes with and without photophoresis in the system hLlG (HR 4796A).
The particles start on circular orbits with $r_0 = 5 \, \AU$. In fact, initial circularity
is not  important, because elliptic orbits are quickly circularized by the gas drag force.
The initial value $r_0$ does not matter either, because
the particles migrate towards their equilibrium orbits within several thousand years.
While small, outward migrating bodies travel slowly towards their equilibrium distance
$\stabilitydist$, approaching it asymptotically, the larger, inward migrating ones almost 
`drop' onto their stability orbits, being stopped almost instantaneously.
Note, however, that `instantaneous' here
refers to a deceleration process taking several centuries. Therefore, the large particles
travel essentially on Keplerian orbits, corresponding to the second case
described in sect. 4.
With increasing particle size, though, this process takes longer, as larger objects experience 
weaker drag acceleration than smaller ones.
Another conclusion from Fig. \ref{rt_bsp_hr} is that the stability radii are pushed outward by 
photophoresis.

\subsection{Equilibrium distance}
\label{eq_distance}

We now calculate the equilibrium distance which, according to Eq.~(\ref{vau_r}),
must satisfy
\formula
\beta + \chi - \eta = 0 \, .
\label{summe}
\formulaend
From this, an implicit expression for the radius $\stabilitydist$ of the stable orbit can be derived:
\formula
r_\mathrm{stab}(s)
= \frac
  {v_K^2(r_\mathrm{stab}) \, (1 - \eta(r_\mathrm{stab}))}
  {a_\mathrm{grav, \, eff}(r_\mathrm{stab}, \, s) - a_\mathrm{phot}(r_\mathrm{stab}, \, s)} \, .
\label{r_stab}
\formulaend

\begin{figure}[h]
\centerline{
\epsfxsize = 0.95\columnwidth
\epsffile{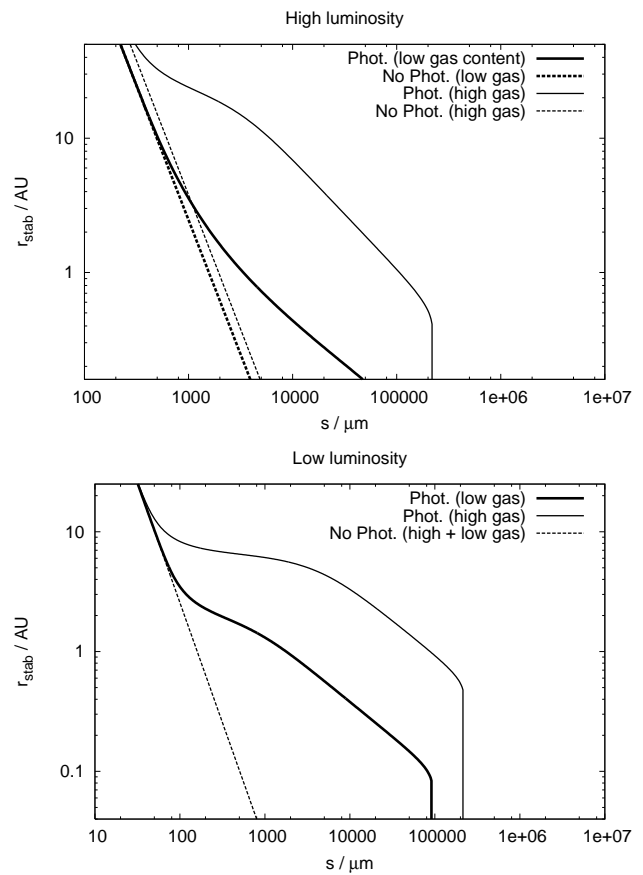}}
\caption{\label{r_stab_plot} Stability radii as functions of particle radius $s$ for
two high-luminosity systems (top) and two low-luminosity ones (bottom).
In the bottom panel, the `No Photophoresis' curve is
identical for both systems, as they differ only in the value of gas density,
which does not affect the stability condition if photophoresis is not taken
into account ($\eta$ only depends on the density exponent $q$, not its
absolute value -- see Eq. \ref{eta_lang}). Again, the lower borders
of the plots correspond to the evaporation limits.}
\end{figure}

\begin{figure}[h]
\centerline{
\epsfxsize = 0.95\columnwidth
\epsffile{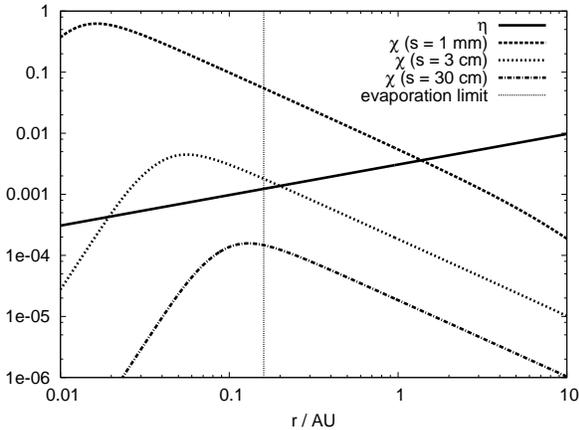}}
\caption{\label{nosolution} The force ratios $\eta$ and $\chi$ as functions of $r$
for different particle sizes, in the hLlG system (HR 4796A).
The number of intersections between the curves
corresponds to the number of solutions of (\ref{summe})--(\ref{r_stab}).
}
\end{figure}

The equilibrium distance $\stabilitydist$ depends on the system's parameters as well as
on the particle size.
Fig. \ref{r_stab_plot} shows solutions of Eq. (\ref{r_stab}), computed with and without 
photophoresis.
It can be clearly seen that
photophoresis significantly increases the radii of stable orbits ~-- as expected:
photophoresis pushes particles away from the star.
The curves exhibit a characteristic shape: after branching off from the
solutions computed without photophoresis, they flatten.
While for hLhG there is only a small region in which the curve is flatter,
the two low-luminosity systems have well-developed plateaus,
the length and distance from the star of which increase with gas density.
The reason for this is explained below. Note also, that
the plateaus in the $\stabilitydist$-curve correspond to areas,
where solids have a higher surface density. At a
certain particle size, the curves drop to zero suddenly. This is a
consequence of mean free path becoming smaller than particle sizes --
in this area, Eq. (\ref{phot_beschl}) is not a good approximation anymore,
and Eq. (\ref{phot_beschl_allkn}) should be used instead to calculate the
photophoretic acceleration. Because in the low Knudsen number regime
photophoresis decreases in strength as body size increases, Eq.
(\ref{summe}) and (\ref{r_stab}) do not have solutions above a certain
body radius, which causes $r_\mathrm{stab}(s)$ to drop to zero, so that
there are no stable orbits for solids above this critical size. These
considerations are visualized in Fig. \ref{nosolution}, where $\eta$
and $\chi$ are plotted as functions of $r$ for different radii $s$. 
The third force parameter,  $\beta$, can be neglected for
large particles ($s \gg 100 \, \micron$), and therefore is not shown.
We can identify two cases.
For small bodies, the $\chi(r)$-curve
has two intersections with $\eta(r)$. This means that in a certain size range,
(\ref{summe})--(\ref{r_stab}) have got two solutions. In our work we consider
only the one farther out from the star, since in transitional disks we expect most
particles to drift inward from larger radii, where they are produced by
collisions between left-over planetesimals in exo-Kuiper belts.
Besides, for three out of four systems (hGlL is the exception), the
inner solution lies in the sublimation zone.
As $s$ increases further, the $\chi(r)$-curve moves downward 
until it no longer crosses the $\eta$-curve, and thus no
solution of (\ref{summe})--(\ref{r_stab}) remains. This means that above a
certain size no stable orbits exist.

\begin{figure}[h]
\centerline{
\epsfxsize = 0.95\columnwidth
\epsffile{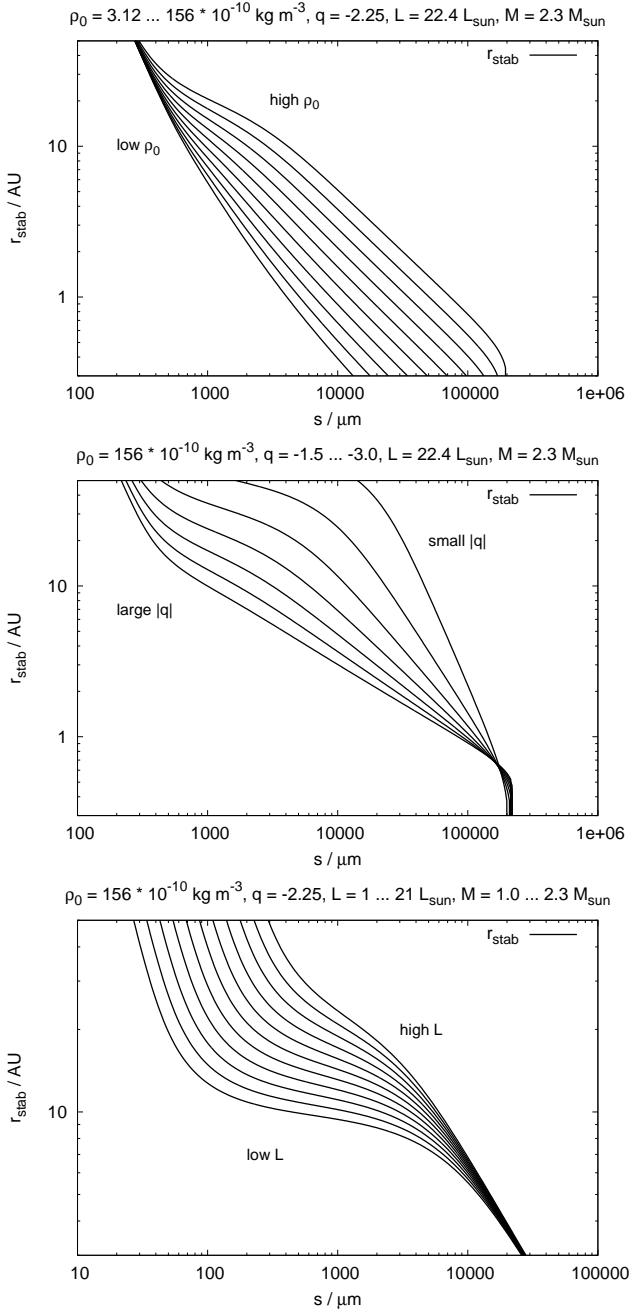}}
\caption{\label{multi} Dependence of the equilibrium distance $\stabilitydist$ on
the gas density $\rho_0$,
the density exponent $q$,
and stellar luminosity $L_\star$.
Top: $\rho_0$ running, $q$ and $L_\star$ fixed.
Middle: $q$ running, $\rho_0$ and $L_\star$ fixed.
Bottom: $L_\star$  running, $q$ and $\rho_0$ fixed.
The parameter values are indicated on top of each panel.}
\end{figure}

Next, we determine how the exact shape of $\stabilitydist$ is influenced by the
choice of system parameters. Therefore, in Fig. \ref{multi}, we plot the
stability radii for different choices of $L_\star$, $\rho_0$ and $q$.
Note that, when varying $L_\star$ in the bottom panel of Fig. \ref{multi},
we change $M_\star$ accordingly.
Analyzing the plots in Fig. \ref{multi}, we can identify the following dependencies:
\begin{itemize}
\item Higher {\em gas density} $\rho_0$ moves the curves' middle parts to larger $r$, which is, 
of course, a consequence of photophoresis being directly dependent on gas pressure.
Also, the limiting particle size beyond which no
stable orbits exist depends almost exclusively on $\rho_0$, while the influence
of the other parameters is rather weak.
\item {\em The density exponent} $q$, which determines the slope of
$\rho(r)$, also controls the slope of $\stabilitydist$: a lower absolute value of
$q$ produces a steeper curve, as the gas disk spreads to a
greater distance from the star.
\item {\em Stellar luminosity} $L_\star$ determines the characteristic
shape of the $\stabilitydist$ curve. While large luminosities produce a monotonically
decreasing curve, lower (\ie solar) ones generate a plateau at a certain
distance from the star, which gives rise to a concentration of particles of
different sizes in that region.
\end{itemize}

The mechanism behind the third effect is to be found in the denominator of
Eq. (\ref{phot_beschl}). As we have seen in Fig. \ref{comp}, in different size
ranges three different processes reduce the photophoretic effect -- heat
conduction in small particles, radiation for larger ones and heat conduction
into the surrounding gas for very large objects. If the stellar luminosity is
low, radiation becomes important only for relatively large sizes, while for
smaller objects only $\Terma$ is relevant in Eq. (\ref{phot_beschl}). As
$\Terma$ is \textit{independent of size}, the resulting curve tends to run
parallel to the $s$-axis, creating the plateau.

In the following, we
formulate these considerations quantitatively.
We are going to compute solutions of Eq. (\ref{r_stab}) for very large and very small
particles, using the simplified large-Knudsen approximation for the
photophoretic acceleration given in Eq. (\ref{phot_beschl}). We need to calculate the following ratios:
\formula
\chi(\Terma) = \frac{a_\mathrm{phot, \, \Terma}}{a_\mathrm{grav}}
\formulaend
and
\formula
\chi(\Termb) = \frac{a_\mathrm{phot, \, \Termb}}{a_\mathrm{grav}} \, ,
\formulaend
where $a_\mathrm{phot, \, \Terma}$ and $a_\mathrm{phot, \, \Termb}$ denote the
accelerations due to
(large Knudsen regime) photophoretic 
force using exclusively $\Terma$ and $\Termb$ in the denominator, respectively. The first case
is a good approximation for small particle radii ($s \le 1 \, \mm$, compare Fig. 
\ref{comp}), the second one for larger, but sub-meter objects ($1 \, \mm \le s \le 1 \, \meter$).

The quantities
$\chi(\Terma)$ and $\chi(\Termb)$ can be computed from the following expressions:
\begin{eqnarray}
\chi(\Terma)
&=&
1.0 \times 10^8 \,
\units{M_\star}{M_\odot}^{-1} \,
\units{L_\star}{L_\odot} \,
\units{\mu_\mathrm{g}}{2.34}^{-1} \,
\nonumber \\
&\times& \,
\units{\rho_\mathrm{d}}{1000 \, \kilogram \, \meter^{-3}}^{-1} \,
\rho_0 \, \units{r}{1 \, \AU}^q
\nonumber \\
&\equiv&
\factorchia \; \units{r}{1 \, \AU}^q 
\label{chi_terma}
\end{eqnarray}
and
\newpage
\begin{eqnarray}
\chi(\Termb)
&=&
2.1 \times 10^{10} \,
\units{M_\star}{M_\odot}^{-1} \, 
\units{L_\star}{L_\odot}^{1/4} \,
\units{\mu_\mathrm{g}}{2.34}^{-1} \,
\nonumber \\
&\times& \,
\units{\rho_\mathrm{d}}{1000 \, \kilogram \, \meter^{-3}}^{-1} \,
\rho_0 \, \units{r}{1 \, \AU}^{q+3/2} \,
\units{s}{1 \, \micron}^{-1} \,
\nonumber \\
&\equiv& 
\factorchib \; \units{r}{1 \, \AU}^{q+3/2} \, \units{s}{1 \, \micron}^{-1} .
\label{chi_termb}
\end{eqnarray}

It is easy to show that in the first (small-particle) case, Eq.
(\ref{r_stab}) can be solved for $s$:
\formula
\frac{s_\mathrm{stab, \, small}(r)}{1 \, \micron}
= \frac{\factorbeta}{\factoreta \, \, (r \, [\AU])^{1/2} - \factorchia \, \, (r \, [\AU])^{q}} \, .
\label{ssmall}
\formulaend
For definition of $\factorbeta$ and $\factoreta$, see Eqs. (\ref{beta}) and
(\ref{eta_lang}).
This allows us to calculate the size of particles on stable orbits at a certain stellar 
distance $r$. Since $q$ is negative, $s_\mathrm{stab, \, small}$ tends to infinity as $r$ 
approaches zero. This produces the curve's  plateau, \ie the concentration belt.
From Eq. (\ref{ssmall}), the belt's radius computes to:
\formula
r_\mathrm{belt} = \left( \frac{\factoreta}{\factorchia} \right)^{1/(q-1/2)}
\label{r_outer}
\formulaend
For the four systems considered, the resulting distances are listed
in Table~\ref{tab_outer_belt}.

\begin{table}[h!]
\caption{\label{tab_outer_belt} Existence and parameters of the belt.}
\begin{tabular}{lrl}
\hline
System                & $r_\mathrm{belt} / \AU$  & $C$ \\
\hline
lLlG (`older' SN)     &  2.0                      & 0.8 \\  
lLhG (`younger' SN)   &  {\bf 6.5}                & {\bf 2.6} \\
hLlG (HR 4696A)       &  4.0                      & 0.1 \\
hLhG (HD 141569A)     & 22.8                      & 0.07\\
\hline
\end{tabular}
\end{table}

In the second (large-particle) case, the corresponding expression
\formula
\frac{s_\mathrm{stab, \, large}(r)}{1 \, \micron}
=
\frac{\factorbeta \, \, (r \, [\AU])^{-1/2} \, + \, \factorchib \, \, (r \, 
[\AU])^{q+1}}{\factoreta}
\label{slarge}
\formulaend
can be derived.

\begin{figure}[t]
\centerline{
\epsfxsize = 0.95\columnwidth
\epsffile{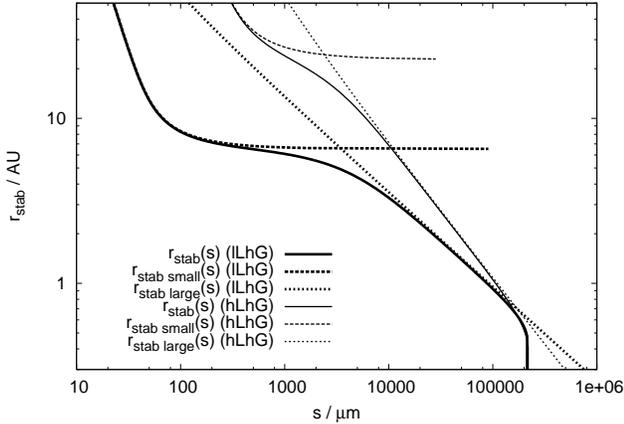}}
\caption{\label{approximation} `Exact' function $r_\mathrm{stab}$ (solid lines)
and its two approximations
(dashed lines: Eq.~\ref{ssmall}; dotted: Eq.~\ref{slarge}),
for the lLhG (thick lines) and the hLhG (thin) systems. Note how the behavior
of $r_\mathrm{stab, \, small}$ produces the curve's plateau -- in the higher
luminosity case, $r_\mathrm{stab, \, large}$ `takes over' at smaller
particle sizes, thus keeping $r_\mathrm{stab, \, small}$ from producing the
effect.}
\end{figure}

The condition for the formation of the belt is that at its position, the value of 
$s_\mathrm{stab, \, large}$ has to be sufficiently larger than $s_\mathrm{stab, \,
small}$. If that is the case, the asymptotic behavior of $s_\mathrm{stab,
\, small}$ is seen in the `exact' curve too, otherwise it is overridden by
$s_\mathrm{stab, \, large}$ (see Fig. \ref{approximation}).
To evaluate this condition numerically, we set
\formula
r_\mathrm{belt}' = r_\mathrm{belt} \, + \, \Delta r
\formulaend
where for $\Delta r$ a sufficiently small value has to be chosen. We use
$\Delta r = 0.1 \, \AU$.
This is necessary because $s_\mathrm{stab, \, small}$ 
cannot be computed at $r_\mathrm{belt}$, as the denominator of Eq.
(\ref{ssmall}) becomes zero at this point.
Then, the ratio of the two approximating functions at $r_\mathrm{belt}'$ can
be used as a measure of how pronounced the belt is:
\formula
C \equiv s_\mathrm{stab, \, large}(r_\mathrm{belt}') \, / \, s_\mathrm{stab, \,
small}(r_\mathrm{belt}') .
\formulaend
The values of $C$ are listed in Table~\ref{tab_outer_belt}.
Comparing with Fig. \ref{r_stab_plot}, we find that for $C \ll 1$, the
plateau does not appear at all, or is only marginal. For $C \approx
1$, the belt is well developed. The particle size range it encompasses
and its degree of concentration (\ie\ the 'flatness' of the curve in that area)
increase with $C$, which makes it a direct measure for the system's tendency
to produce a belt.
In fact, other choices for $\Delta r$ are possible -- for them, the values
with which $C$ has to be compared, have to be changed. For $\Delta r = 0.1 \,
\AU$, the critical value above which well-formed belts appear is $C = 1$.

We see that low luminosities and high gas densities are crucial for belt
formation.
Since $\Termb \propto T^4 \propto L_\star$, with rising luminosity $\Termb$
becomes important at smaller particle sizes, superseding $\Terma$ before the
plateau of $s_\mathrm{stab, \, small}$ is reached. Because they increase the overall strength of
photophoresis, higher gas densities push the plateau outward, reducing its slope
-- and thus increasing the degree of particle concentration in the belt.

Fig. \ref{approximation} demonstrates how the exact solution of
Eq. (\ref{r_stab}) is approximated by Eq. (\ref{ssmall}) and Eq. (\ref{slarge})
in different 
particle size ranges for the lLhG and hLhG models.

\subsection {Particle size range}

\begin{figure}[h]
\centerline{
\epsfxsize = 0.95\columnwidth
\epsffile{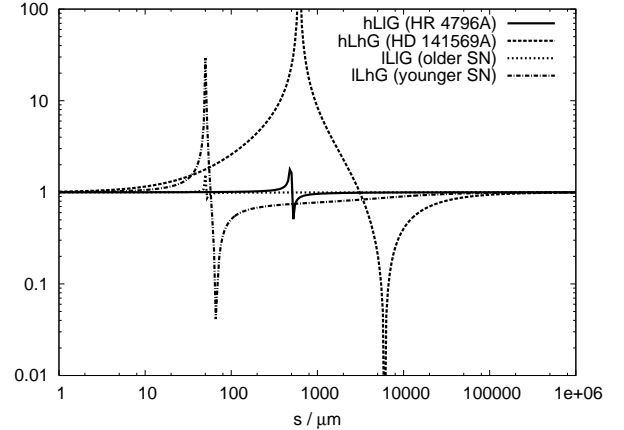}}
\caption{\label{ratio} Ratio of radial velocities, computed with and without
photophoresis. The stronger the deviation from unity, the more important is photophoresis.
The stellar distance is $r = 10 \, \AU$}
\end{figure}

As we have seen, particles of different sizes are influenced by
photophoresis to a different extent.
Its relative importance depends on their particle radius, declining for very
small (micrometer-range) and very large (above meter-range) objects. 
In between, photophoresis plays an important role. In order to analyze the
critical size range, we plot the ratio
$x \equiv | v_\mathrm{r, \, phot} / v_\mathrm{r, \, no \, phot} |$
as a function of $s$ in Fig. \ref{ratio}, where $v_\mathrm{r, \, phot}$ and 
$v_\mathrm{r, \, no \, phot}$ are the radial velocity with and without photophoresis
respectively. The distance is set to $r = 10 \, \AU$.
Note that the interval between the two peaks corresponds to the size range
in which the two $v_\mathrm{r}$ have got different signs, because the
stability radius with photophoresis is greater than 10 AU, while the one
calculated without photophoresis is smaller.

If we use the criterion that $|x - 1| \ge 0.001$, photophoresis has to be
taken into account in the size ranges listed in Table~\ref{tab_size_range}.
Note, however, that for $s \le 10 \, \micron$, the above evaluation is no
longer necessarily valid, as for small particles, the photophoretic force
can reverse its direction \citep{tehranian_etal_2001}.
Also, for hLhG (HD 141569A), $\beta(s_\mathrm{min})$ is greater than 0.5, thus
these particles are $\beta$-meteoroids which escape from the system.

\begin{table}[h!]
\caption{\label{tab_size_range} Size range of particles affected by photophoresis.}
\begin{tabular}{lrl}
\hline
System                &  $s_\mathrm{min}$ & $s_\mathrm{max}$ \\
\hline
lLlG (`older' SN)     &  $9 \, \micron$  & $3 \, \cm$ \\
lLhG (`younger' SN)   &  $1 \, \micron$  & $1.5 \, \meter$ \\
hLlG (HR 4696A)       & $10 \, \micron$  & $3.2 \, \cm$ \\
hLhG (HD 141569A)     &  $1 \, \micron$  & $6 \, \meter$\\
\hline
\end{tabular}
\end{table}

%-----------------------------------------------------------------------------------------
\section{Rotation of particles}
%-----------------------------------------------------------------------------------------

Until now, we considered only nonrotating particles. We have to check
whether this assumption is realistic, since the rotation of bodies can
transport thermal energy from the dark to the lit side, thus
banishing the photophoretic effect. There is a number of different
mechanisms that can induce particle rotation:
\begin{itemize}
\item {\em Collisionally induced rotation.} During the formation of
bodies as well as during their life, they collide with other particles,
which can change their orbits as well as transfer angular momentum to them,
spinning them up.
\item {\em Rotation induced by gas drag or radiation forces.} If
particles are not exactly spherical, any force acting on their surface will change
their angular momentum.
For instance, radiation pressure may spin-up the particles (the so-called
windmill effect, \citeauthor{paddack-rhee-1975} \citeyear{paddack-rhee-1975}),
but may also stabilize/align then \citep{draine-weingartner-1996}.
Such effects are beyond the scope of this paper.
\end{itemize}
Whether rotation is able to subdue photophoresis depends on four different timescales:
$t_\mathrm{coll}$, the typical time between two collisions;
$t_\mathrm{s}$, the gas coupling time which also determines the time
needed to slow down rotation;
$t_\mathrm{heat}$, the thermal
relaxation time needed to establish a stable heat gradient within the particle;
and $t_\mathrm{rot}$, the typical rotation period.

The collisional time $t_\mathrm{coll}$ can be estimated in a standard way:
\formula
t_\mathrm{coll} = \frac{1}{n_\mathrm{p} v \sigma_\mathrm{coll}} \, ,
\formulaend
where $n_\mathrm{p}$ is the particle number density, $v$ the collisional
velocity and $\sigma_\mathrm{coll} = \pi \, (s_1 + s_2)^2$
is the collisional cross section
for spherical particles of radii $s_1$ and $s_2$.
With
\formula
n_\mathrm{p} = \frac{\rho_\mathrm{d}}{\rho_\mathrm{bulk} \, \frac{4}{3} \,
\pi s^3} \, ,
\formulaend
and the standard assumption $\rho_\mathrm{d} / \rho_\mathrm{g} = 10^{-2}$, 
we get for particles of equal size \citep[\mbox{\em cf.}][their Eq. 15]{krausswurm_06}:
\formula
t_\mathrm{coll} = 33 \, \frac{\rho_\mathrm{bulk} s}{\rho_\mathrm{g} v} \, .
\label{tcoll}
\formulaend
The stopping time $t_\mathrm{s}$ was computed  in section \ref{gasdrag}, Eq.
(\ref{T_s}). Finally, the thermal relaxation time $t_\mathrm{heat}$ is
given by \citep[see][Eq. 6]{krausswurm}:
\formula
t_\mathrm{heat} = \frac{\rho_\mathrm{bulk} c_\mathrm{d} s^2}{k_\mathrm{th}}
\, ,
\label{theat}
\formulaend
where $c_\mathrm{d} \approx 1000 \, \watt \meter^{-1} \kelvin^{-1}$ is the
particle's heat capacity.

We can use two different approaches to deal with rotation. One is to
compare collision time to stopping time: if $t_\mathrm{coll} / t_\mathrm{s}
\gg 1$, damping of rotation occurs faster than excitation by impacts, and therefore
photophoresis is not significantly weakened. Using Eqs. (\ref{tcoll}) and
(\ref{T_s}), we get
\formula
\frac{t_\mathrm{coll}}{t_\mathrm{s}} = 25 \, \frac{v_\mathrm{th}}{v} \, .
\formulaend
Another option is to check whether rotation,
once it has been excited by collisions, is sufficiently slow to allow
a stable heat gradient to be established within the particle: this
is the case if $t_\mathrm{rot} / t_\mathrm{heat} \gg 1$. In order to find a
rough estimate for $t_\mathrm{rot}$, we assume that the particle transforms the
entire kinetic energy of its radial motion $v_\mathrm{r}$ into rotational
energy:
\formula
\bytwo{1} \, \left( \frac{2 \, \pi}{t_\mathrm{rot}} \right)^2 \,
J_\mathrm{sphere} = \bytwo{1} \, m \, v_\mathrm{r}^2 \, ,
\formulaend
with the solid sphere's moment of inertia $J_\mathrm{sphere} = (2/5) m s^2$, we get
\formula
t_\mathrm{rot} = \sqrt{\frac{8}{5}} \, \frac{\pi s}{v_\mathrm{r}} \, .
\formulaend

\begin{figure}[h]
\centerline{
\epsfxsize = 0.95\columnwidth
\epsffile{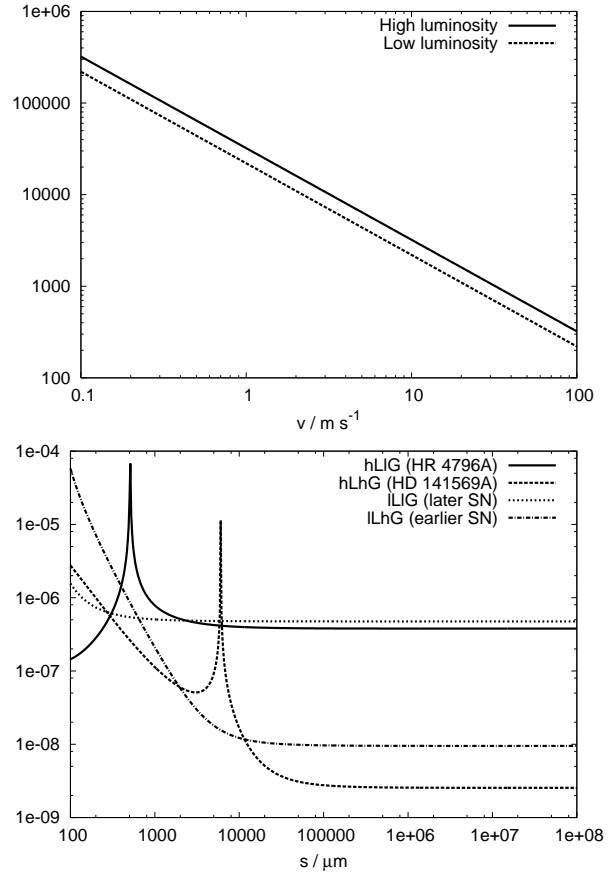}}
\caption{\label{rotation}
Ratios of typical timescales.
Top: $t_\mathrm{coll} / t_\mathrm{s}$ as a function of $v$.
Bottom: $t_\mathrm{rot} / t_\mathrm{heat}$ as a function of $s$.}
\end{figure}

Fig. \ref{rotation} shows $t_\mathrm{coll} / t_\mathrm{s}$ as a function of
$v$ and $t_\mathrm{rot} / t_\mathrm{heat}$ as a function of $s$.
The distance is chosen to be $r = 10 \, \AU$.
We see that collisional timescales are longer than stopping timescales by
many decades, thus allowing rotation to decline between collisions.
Nonetheless, rotation frequencies can be high enough to cancel out stable
temperature gradients.
We conclude that photophoresis can be suppressed
directly after collisions through rapid rotation, but remains effective for 
most of the time because gas drag damps spin faster than collisions can
excite it.

%-----------------------------------------------------------------------------------------
\section{Conclusions and discussion}
%-----------------------------------------------------------------------------------------

In this paper, we have analyzed the effect of photophoresis on the dynamics
of solid particles in the optically-thin, yet sufficiently gas-rich, transitional
disks around young stars.
To this end, we employed a single-body approach and added the photophoretic force
to the standard array of perturbing forces (stellar gravity, direct radiation
pressure, and gas drag). Particle-particle interactions as well as particle growth
were not included.
We find that photophoresis may lead to noticeable corrections to the results
obtained with models that do not take it into account.

Our main results can be summarized as follows:
\begin{enumerate}
\item
Both with and without photophoresis,
solid objects migrate inward or outward, until they reach the stability distance
$\stabilitydist$, where $s$ is the particle radius.
At that distance, radial forces cancel each other in the
particle's own inertial system, and the orbital velocity is equal to that of the
gas. The stability distance is a decreasing function of $s$, therefore particles are
sorted according to size, with larger bodies accumulating closer to the star.
These results fully  agree with those by \cite{tak_arty}.
\item
Photophoresis increases the stability radii, moving objects to
larger radial distances.
The effect is noticeable in the size range from several micrometers to
several centimeters (for older transitional disks) or even several meters
(for younger, more gaseous, ones).
\item
The steady-state distribution of solids is completely characterized by
the function $\stabilitydist$, the shape of which depends on the system's
parameters:
  \begin{itemize}
  \item Higher gas densities move the curve to larger distances
  without changing its overall shape. Gas density also controls the maximum 
  particle size up to which stable orbits can exist.
  \item
  The steepness of the gas density radial profile determines
  the slope of $\stabilitydist$, with flatter profiles generating steeper
  curves.
  \item
  Stellar luminosity determines the curve's shape. While high
  luminosities ($\sim 20 \, L_\odot$) produce simple decreasing curves, low
  (solar) luminosities generate a plateau at a distance
  from the star that can be computed with the aid of Eq. (\ref{r_outer}).
  In this area, objects in a certain size range may accumulate.
  \end{itemize}
\item
Particle rotation tends to reduce the photophoretic effect. Our estimates suggest, however,
that it is damped by gas drag quickly enough to keep photophoresis at work.
\end{enumerate}

Our work predicts the formation of a particle concentration belt at
a certain distance from the star. For the high luminosity systems, it is
not very pronounced (hLhG) or does not appear at all (hLlG).
Furthermore, the radii of the observed rings around HD 141569A are an order
of magnitude larger than that of the slight concentration belt predicted for
the hLhG system. It is not likely
therefore that the observed structures around HR 4796A and HD 141569A are caused by
photophoresis.
While photophoresis is probably active in transitional
disks, the circumstellar rings of HR 4796A and HD 141569 must be shaped by
other forces and effects, such as gravitational sculpting by planets or interactions
with stellar companions.
Alternatively, a rapid decline of gas density at the disks' edges \citep[][]{tak_arty} 
or a recently proposed dust-gas instability \citep[][]{besla_wu_07} may cause
particles to accumulate there.

The model presented here is rather exploratory and rests on a number of simplifying 
assumptions. In the future, we plan to investigate the
problem a more realistic way, lifting some of the assumptions we made
to keep the problem tractable. First, we plan to deal with particle-particle
interactions, taking collisions and growth into account. This can be done in
the style of \cite{krausswurm_06}, increasing the radius through
an exponential ansatz $s = s_0 \, \exp(t/t_0)$. The latter corresponds to the
assumption that the object moving through the nebula collects smaller particles
on its surface.
A more detailed
approach will employ statistical methods \citep{krivov-et-al-2005,krivov-et-al-2006}.
Then, we wish to explore rotation in greater detail, calculating collision
timescales and rotation frequencies using models for two-body collisions.
Also, the effects of gas drag and radiation forces on non-spherical
objects need to be taken into account \citep[see e.g.][]{xu_etal_99}.
Other issues include the variation of physical parameters (density, thermal
conductivity, emissivity, $J_1$) with size and distance from the star, as
well as the global evolution and clearing of dust in the system which defines the
time around which photophoresis can come into play.

In spite of these unknowns, we have demonstrated that photophoretic force
in transitional circumstellar disks cannot be neglected, and
has to be included in elaborate models of such systems.

%-----------------------------------------------------------------------------------------
%-----------------------------------------------------------------------------------------
\newpage
\textit{Acknowledgments.} We wish to thank Gerhard Wurm for useful
discussions, J\"urgen Blum and Bo Gustafson for pointing out
several papers on the photophoretic effect and the anonymous referee for
helpful comments. F.H. is supported by the 
graduate student fellowship of the Thuringia State.

%-----------------------------------------------------------------------------------------
%-----------------------------------------------------------------------------------------

\input 8322jour

\input 8322bibl
\end{document}

%% file: 8322jour.tex
\newcommand{\AAp}      {Astron. Astrophys.}
\newcommand{\AApSS}    {AApSS}
\newcommand{\AApT}     {Astron. Astrophys. Trans.}
\newcommand{\AdvSR}    {Adv. Space Res.}
\newcommand{\AJ}       {Astron. J.}
\newcommand{\AN}       {Astron. Nachr.}
\newcommand{\ApJ}      {Astrophys. J.}
\newcommand{\ApJS}     {Astrophys. J. Suppl.}
\newcommand{\ApSS}     {Astrophys. Space Sci.}
\newcommand{\ARAA}     {Ann. Rev. Astron. Astrophys.}
\newcommand{\ARevEPS}  {Ann. Rev. Earth Planet. Sci.}
\newcommand{\BAAS}     {BAAS}
\newcommand{\CelMech}  {Celest. Mech. Dynam. Astron.}
\newcommand{\EMP}      {Earth, Moon and Planets}
\newcommand{\EPS}      {Earth, Planets and Space}
\newcommand{\GRL}      {Geophys. Res. Lett.}
\newcommand{\JGR}      {J. Geophys. Res.}
\newcommand{\MNRAS}    {MNRAS}
\newcommand{\PASJ}     {PASJ}
\newcommand{\PASP}     {PASP}
\newcommand{\PSS}      {Planet. Space Sci.}
\newcommand{\SolPhys}  {Sol. Phys.}
\newcommand{\SolSysRes}{Sol. Sys. Res.}
\newcommand{\SSR}      {Space Sci. Rev.}